\documentclass[conference]{Components/Metafiles/IEEEtran}
\IEEEoverridecommandlockouts
\usepackage{graphicx}
\usepackage{pgf}
\usepackage{tikz}
\usetikzlibrary{arrows}
\usetikzlibrary{arrows,automata}
\usetikzlibrary{spy}
\usetikzlibrary{patterns}
\usepackage[latin1]{inputenc}
\usepackage{subcaption}
\usepackage{amssymb}
\usepackage{tabularx,booktabs}
\usepackage{pgfplots}
\usepackage{lipsum}
\usepackage{dsfont}
\usepackage{tikz-layers}
\usepackage{setspace}
\usepackage{subcaption}
\usepackage[ruled,vlined]{algorithm2e}
\pgfplotsset{compat=1.6}
\usepgfplotslibrary{statistics}
\usepgfplotslibrary{colorbrewer} %added
\pgfplotsset{compat = 1.15, cycle list/Set1-8} %added
\usetikzlibrary{pgfplots.statistics, pgfplots.colorbrewer} %added
\usepackage{pgfplotstable}
\usepackage{filecontents}%added
\usepackage{amsmath}
\usepackage{bm}
\usepackage{mathtools}
\usepackage{cleveref}
\usepackage[acronym,toc]{glossaries}
\usepackage[cmintegrals]{newtxmath}
\usepackage{stackengine}
\usepackage{afterpage}
\usepackage[euler]{textgreek}
\usepackage{lgreek}
\usepackage[hyphens]{url}

\def\BibTeX{{\rm B\kern-.05em{\sc i\kern-.025em b}\kern-.08em
    T\kern-.1667em\lower.7ex\hbox{E}\kern-.125emX}}

\newcommand{\bc}{\mbox {\boldmath $c$}}

\newcommand{\bw}{\mbox {\boldmath $w$}}

\newcommand{\calU}{\mathcal{U}}

\newcommand{\calX}{\mathcal{X}}

\newcommand{\R}{\mathbb{R}}

\definecolor{mycolor1}{rgb}{0.29, 0.59, 0.82}%
\definecolor{mycolor2}{rgb}{1.0, 0.4, 0.6}
\definecolor{mycolor3}{rgb}{0.92900,0.69400,0.12500}%
\definecolor{mycolor4}{rgb}{0.71,0.49,0.86}
\definecolor{mycolor5}{rgb}{0.12, 0.3, 0.17}
\definecolor{mycolor6}{rgb}{0.43, 0.21, 0.1}
\definecolor{mycolor7}{rgb}{0.52, 0.73, 0.4}
\definecolor{mycolor8}{rgb}{0.98, 0.38, 0.5}
\definecolor{vColor}{rgb}{0.12, 0.3, 0.17}

\makeglossaries

\newacronym{TV}{TV}{television}
\newacronym{DTTB}{DTTB}{digital television terrestrial broadcasting}
\newacronym{DVB}{DVB}{Digital Video Broadcast}
\newacronym{DVB-H}{DVB-H}{Digital Video Broadcast-Handheld}
\newacronym{ATSC}{ATSC}{Advanced Television System Committee}
\newacronym{ATSC-M/H}{ATSC-M/H}{Advanced Television System Committee - Mobile/Handheld}
\newacronym{IPTV}{IPTV}{Internet Protocol television}
\newacronym{IP}{IP}{Internet Protocol}

\newacronym{UE}{UE}{user equipment}
\newacronym{PC}{PC}{personal computer}
\newacronym{RAN}{RAN}{radio access network}
\newacronym{CN}{CN}{core network}
\newacronym{MS}{MS}{mobile station}

\newacronym{ITU-R}{ITU-R}{International Telecommunications Union - Radiocommunication Sector}
\newacronym{IMT-Advanced}{IMT-Advanced}{International Mobile Telecommunications Advanced}
\newacronym{4G}{4G}{fourth-generation of mobile phone communications and Internet access technology}

\newacronym{3gpp}{3GPP}{3rd Generation Partnership Project}
\newacronym{GSM}{GSM}{Global System for Mobile Communications}
\newacronym{UMTS}{UMTS}{Universal Mobile Telecommunications System}
\newacronym{HSPA}{HSPA}{High Speed Packet Access}
\newacronym{lte}{LTE}{Long-Term Evolution}
\newacronym{lte-a}{LTE-A}{Long-Term Evolution Advanced}

\newacronym{e-UTRAN}{e-UTRAN}{evolved Universal Terrestrial Radio Access Network}
\newacronym{eNB}{eNB}{e-UTRAN NodeB}
\newacronym{gNB}{gNB}{gNodeB}
\newacronym{EPC}{EPC}{Evolved Packet Core}

\newacronym{MBMS}{MBMS}{Multimedia and Broadcast Multicast Service}
\newacronym{eMBMS}{eMBMS}{Evolved MBMS}
\newacronym{SFN}{SFN}{single-frequency network}
\newacronym{MBSFN}{MBSFN}{MBMS single-frequency network}
\newacronym{BM-SC}{BM-SC}{Broadcast/Multicast Service Center}
\newacronym{MBMS GW}{MBMS GW}{MBMS Gateway}
\newacronym{MME}{MME}{Mobility Management Entity}
\newacronym{MCE}{MCE}{Multi-cell/multicast Coordinating Entity}
\newacronym{SYNC}{SYNC}{synchronization}
\newacronym{MCCH}{MCCH}{Multicast Control Channel}
\newacronym{MTCH}{MTCH}{Multicast Traffic Channel}
\newacronym{MCH}{MCH}{Multicast Channel}
\newacronym{PMCH}{PMCH}{Physical Multicast Channel}
\newacronym{PDSCH}{PDSCH}{Physical Downlink Shared Channel}

\newacronym{IEEE}{IEEE}{Institute of Electrical and Electronics Engineers}
\newacronym{WiMAX}{WiMAX}{Worldwide Interoperability for Microwave Access}
\newacronym{ASN}{ASN}{access service network}
\newacronym{ASN-GW}{ASN-GW}{ASN gateway}
\newacronym{CSN}{CSN}{Connectivity Service Network}

\newacronym{PA}{PA}{power amplifier}

\newacronym{NI}{NI}{National Instruments}

\newacronym{TDD}{TDD}{time-division duplex}
\newacronym{FDD}{FDD}{frequency-division duplex}
\newacronym{UDP}{UDP}{User Datagram Protocol}
\newacronym{APP}{APP}{application}
\newacronym{mac}{MAC}{medium access control}
\newacronym{phy}{PHY}{physical}
\newacronym{RLC}{RLC}{radio link control}
\newacronym{FIFO}{FIFO}{first-in first-out}
\newacronym{CRC}{CRC}{cyclic redundancy check}
\newacronym{SAP}{SAP}{service access point}
\newacronym{FEC}{FEC}{forward error correction}
\newacronym{IF}{IF}{intermediate frequency}
\newacronym{RF}{RF}{radio frequency}
\newacronym{mimo}{MIMO}{multiple-input and multiple-output}
\newacronym{MCS}{MCS}{modulation and coding scheme}

\newacronym{SPC}{SPC}{superposition coding}
\newacronym{SVC}{SVC}{Scalable Video Coding}
\newacronym{GM}{GM}{generic multicasting}
\newacronym{SCM}{SCM}{superposition coded multicasting}
\newacronym{SIC}{SIC}{successive interference cancellation}

\newacronym{st}{ST}{secondary transmitter}
\newacronym{pt}{PT}{primary transmitter}
\newacronym{sr}{SR}{secondary receiver}
\newacronym{pr}{PR}{primary receiver}
\newacronym{su}{SU}{secondary user}
\newacronym{pu}{PU}{primary user}

\newacronym{awgn}{AWGN}{additive white Gaussian noise}

\newacronym{pdf}{PDF}{probability density function}
\newacronym{cdf}{CDF}{cumulative density function}
\newacronym{ccdf}{CCDF}{complementary CDF}
\newacronym{iid}{i.i.d.}{independent and identicaly distributed}
\newacronym{rf}{RF}{radio frequency}

\newacronym{dd}{DD}{Device-to-Device}
\newacronym{ddu}{DDU}{Device-to-Device user}
\newacronym{dds}{DDS}{Device-to-Device system}
\newacronym{ddt}{DT}{DDU transmitter}
\newacronym{ddr}{DR}{DDU receiver}

\newacronym{bs}{BS}{base station}
\newacronym{bsu}{BSU}{base station associated user}
\newacronym{bss}{BSS}{base station associated system}
\newacronym{bst}{BT}{BSU transmitter}
\newacronym{bsr}{BR}{BSU receiver}

\newacronym{epg}{EPG}{energy per goodbit}
\newacronym{mepg}{MEPG}{modified energy per goodbit}
\newacronym{ee}{EE}{energy efficiency}
\newacronym{se}{SE}{spectral efficiency}

\newacronym{wrt}{w.r.t.}{with respect to}

\newacronym{kkt}{KKT}{Karush-Kuhn-Tucker}
\newacronym{admm}{ADM}{Alternating Directing Method}
\newacronym{cr}{CR}{cognitive radio}
\newacronym{ssi}{SSI}{soft-sensing information}
\newacronym{csi}{CSI}{channel state information}
\newacronym{qsi}{QSI}{queue state information}
\newacronym{el}{EL}{enhancement layer(s)}
\newacronym{snr}{SNR}{signal-to-noise ratio}

\newacronym{NAL}{NAL}{network abstraction layer}
\newacronym{QP}{QP}{quantization parameter}

\newacronym{ofdm}{OFDM}{orthogonal frequency-division multiplexing}
\newacronym{ofdma}{OFDMA}{orthogonal frequency-division multiple access}
\newacronym{tdma}{TDMA}{time division multiple access}

\newacronym{PUSC}{PUSC}{partial usage of the subchannels}
\newacronym{CFO}{CFO}{carrier frequency offset}
\newacronym{I/Q}{I/Q}{in-phase and quadrature-phase}
\newacronym{ASK}{ASK}{amplitude-shift keying}
\newacronym{PSK}{PSK}{phase-shift keying}
\newacronym{BPSK}{BPSK}{binary phase-shift keying}
\newacronym{QPSK}{QPSK}{quadrature phase-shift keying}
\newacronym{QAM}{QAM}{quadrature amplitude modulation}
\newacronym{PSNR}{PSNR}{peak signal-to-noise ratio}
\newacronym{PELR}{PELR}{packet error and loss rate}

\newacronym{kNN}{\textit{k}-NN}{\textit{k}-nearest neighbor algorithm}
\newacronym{SVM}{SVM}{support vector machines}
\newacronym{nn}{NN}{neural network}
\newacronym{NN}{NN}{neural network}
\newacronym{dnn}{DNN}{deep neural network}
\newacronym{RBF}{RBF}{radial basis function}
\newacronym{RMSE}{RMSE}{root mean squared error}
\newacronym{mse}{MSE}{mean squared error}
\newacronym{lmse}{LMSE}{linear mean square-error estimator}

\newacronym{R2}{$R^2$}{coefficient of determination}

\newacronym{KAUST}{KAUST}{King Abdullah University of Science and Technology}
\newacronym{GSA}{GSA}{Global mobile Suppliers Association}

\newacronym{VoD}{VoD}{video on demand}
\newacronym{HEVC}{HEVC}{High Efficiency of Video Coding}
\newacronym{DASH}{DASH}{Dynamic Adaptive Streaming over HTTP}

\newacronym{PUT}{PUT}{people using television}

\newacronym{ADTVS}{ADTVS}{Audience Driven live TV Scheduling}

\newacronym{arq}{ARQ}{automatic repeat request}

\newacronym{harq}{HARQ}{hybrid automatic repeat request}

\newacronym{sdp}{SDP}{semi-definite programming}

\newacronym{tcp}{TCP}{transmission control protocol}
\newacronym{e2e}{E2E}{end-to-end}
\newacronym{ran}{RAN}{radio access network}
\newacronym{cran}{CRAN}{cloud radio access network}
\newacronym{udcran}{UD-CRAN}{ultra-dense CRAN}
\newacronym{dran}{DRAN}{distributed radio access network}
\newacronym{hcran}{H-CRAN}{hybrid cloud radio access network}
\newacronym{hetnet}{HetNet}{heterogeneous network}
\newacronym{vcran}{V-CRAN}{virtualized CRAN}
\newacronym{ecran}{E-CRAN}{edge-CRAN}
\newacronym{hvcran}{H-VCRAN}{hybrid-virtualized CRAN}

\newacronym{bbu}{BBU}{baseband processing unit}
\newacronym{rrh}{RRH}{remote radio head}
\newacronym{ru}{RU}{radio unit}
\newacronym{rs}{RS}{remote site}
\newacronym{cs}{CS}{central site}

\newacronym{rru}{RRU}{radio resource unit}
\newacronym{rb}{RB}{resource block}
\newacronym{hpn}{HPN}{high-power node}
\newacronym{lpn}{LPN}{low-power node}
\newacronym{mabs}{MaBS}{macro basestation}
\newacronym{ue}{UE}{user equipment}

\newacronym{comp}{CoMP}{coordinated multi-point}
\newacronym{ranaas}{RANaaS}{RAN-as-a-Service}

\newacronym{rof}{RoF}{radio over fiber}
\newacronym{wdm}{WDM}{Wavelength Division Multiplexing}
\newacronym{dls}{DLS}{distributed large scale}

\newacronym{qos}{QoS}{quality of service}
\newacronym{qoe}{QoE}{quality of experience}
\newacronym{qee}{QEE}{quality of energy-efficiency}

\newacronym{gg}{GG}{group-to-group}
\newacronym{ht}{HT}{hyper-transceiver}

\newacronym{fh}{FH}{fronthaul}
\newacronym{dl}{DL}{downlink}
\newacronym{ul}{UL}{uplink}

\newacronym{cp}{CP}{Cell-Processing}
\newacronym{up}{UP}{User-Processing}

\newacronym{co}{CO}{center office}

\newacronym{du}{DU}{digital unit}
\newacronym{lc}{LC}{Line-Card}

\newacronym{onu}{ONU}{optical network unit}
\newacronym{olt}{OLT}{optical line terminal}
\newacronym{osw}{OSW}{optical switch}

\newacronym{es}{ES}{ethernet switch}

\newacronym{ppp}{PPP}{Poisson point process}

\newacronym{mppp}{MPPP}{marked Poisson point process}

\newacronym{sinr}{SINR}{signal to noise and interference ratio}

\newacronym{sir}{SIR}{signal to interference ratio}

\newacronym{mbs}{MBS}{macro basestation}
\newacronym{ap}{AP}{access point}
\newacronym{fap}{FAP}{femto-cell access point}
\newacronym{sap}{SAP}{small-cell access point}
\newacronym{iot}{IoT}{Internet of Things}
\newacronym{ti}{TI}{Tactile Internet}
\newacronym{lsm}{LSM}{linear scalarizing method}

\newacronym{lp}{LP}{Low-Priority}
\newacronym{hp}{HP}{High-Priority}
\newacronym{lpu}{LPU}{Low-Priority user}
\newacronym{hpu}{HPU}{High-Priority user}
\newacronym{lps}{LPS}{Low-Priority system}
\newacronym{hps}{HPS}{High-Priority system}

\newacronym{ttm}{TTM}{time to market}
\newacronym{udn}{UDN}{ultra-dense network}

\newacronym{capex}{CAPEX}{capital expenditure}
\newacronym{opex}{OPEX}{operational expenditure}

\newacronym{cpri}{CPRI}{common public radio interface}
\newacronym{otn}{OTN}{optical transport network}
\newacronym{pon}{PON}{passive optical network}
\newacronym{twdm}{TWDM}{time and wavelength division multiplexing}

\newacronym{ec}{EC}{Edge-Cloud}
\newacronym{cc}{CC}{Central-Cloud}

\newacronym{mmw}{m-Wave}{Milli-Meter wave}

\newacronym{gops}{GOPS}{giga operation per second}
\newacronym{mops}{MOPS}{mega operation per second}

\newacronym{ip}{IP}{internet protocol}
\newacronym{rlc}{RLC}{radio link control}
\newacronym{pdcp}{PDCP}{packet data convergence protocol}

\newacronym{mno}{MNO}{mobile network operator}
\newacronym{prb}{PRB}{physical resource block}
\newacronym{mi}{MI}{modulation index}

\newacronym{wifi}{WiFi}{wireless local area network}
\newacronym{cpu}{CPU}{central processing unit}
\newacronym{vcpu}{VCPU}{virtual CPU}
\newacronym{vm}{VM}{virtual machine}

\newacronym{urs}{UrS}{user requested service}

\newacronym{rsf}{RSF}{radio sub-frame}
\newacronym{siso}{SISO}{single-input single-output}

\newacronym{mec}{MEC}{mobile edge computing}
\newacronym{co2}{CO$_{2}$}{carbo dioxide}

\newacronym{ar}{AR}{augmented reality}
\newacronym{vr}{VR}{virtual reality}
\newacronym{cfp}{CFP}{communication function processing}
\newacronym{ptp}{PTP}{precision time protocol}

\newacronym{voip}{VoIP}{voice over Internet protocol}

\newacronym{da}{DA}{data analytics}

\newacronym{kpi}{KPI}{key performance indicator}

\newacronym{fsmc}{FSMC}{finite state markov chain}
\newacronym{ml}{ML}{machine learning}

\newacronym{5g}{5G}{fifth generation of mobile communication systems}
\newacronym{gnbcu}{gNB-CU}{gNB central unit}
\newacronym{gnbdu}{gNB-DU}{gNB distributed unit}
\newacronym{ecpri}{eCPRI}{common public radio interface}
\newacronym{fl}{FL}{federated learning}
\newacronym{rsrq}{RSRQ}{reference signal received quality}
\newacronym{rsrp}{RSRP}{reference signal received power}
\newacronym{urllc}{URLLC}{ultra-reliable low-latency communications}
\newacronym{embb}{eMBB}{enhanced mobile broadband}

\newacronym{mae}{MAE}{modified autoencoder}
\newacronym{mtc}{MTC}{machine type communication}
\newacronym{mmtc}{mMTC}{massive machine type communication}
\newacronym{pca}{PCA}{principal component analysis}

\newacronym{cps}{CPS}{cyber-physical system}
\newacronym{gnb}{gNB}{gNodeB}
\newacronym{ref}{REF}{reliability enhancement feature}
\newacronym{nfo}{NFO}{network level feature orchestrator}
\newacronym{dc}{DC}{data center}
\newacronym{nf}{NF}{network function}
\newacronym{vnf}{VNF}{virtual network function}
\newacronym{nfv}{NFV}{network functions virtualization}
\newacronym{nssmf}{NSSMF}{network slice subnet management function}
\newacronym{nsmf}{NSMF}{network slice management function}
\newacronym{ai}{AI}{artificial intelligence}
\newacronym{rl}{RL}{reinforcement learning}
\newacronym{ddpg}{DDPG}{deep deterministic policy gradient}
\newacronym{dqn}{DQN}{deep Q-networks}
\newacronym{sac}{SAC}{soft actor-critic}
\newacronym{a2c}{A2C}{advantage actor-critic}
\newacronym{bsac}{BSAC}{branching SAC}
\newacronym{td3}{TD3}{twin delayed deep deterministic policy gradient algorithm}
\newacronym{sgd}{SGD}{stochastic gradient descent}
\newacronym{um}{UM}{unacknowledged mode}
\newacronym{am}{AM}{acknowledged mode}

\begin{document}
\bstctlcite{IEEEexample:BSTcontrol}	
%\title{Trade-offs in Coexistence of Distributed AI and URLLC Traffic}
\title{Interplay between Distributed AI Workflow and URLLC}

\author{
\IEEEauthorblockN{Milad Ganjalizadeh\IEEEauthorrefmark{1}\IEEEauthorrefmark{2}, Hossein S. Ghadikolaei\IEEEauthorrefmark{1}, Johan Haraldson\IEEEauthorrefmark{1}, and Marina Petrova\IEEEauthorrefmark{2}\IEEEauthorrefmark{3}
\thanks{This work was supported by Swedish Foundation for Strategic Research (SSF) under Grant iPhD:ID17-0079.}
}
%\\
% \\
\vspace{3mm}
\IEEEauthorblockA{
% \\
\IEEEauthorrefmark{1}Ericsson Research, Sweden\\
\IEEEauthorrefmark{2}School of Electrical Engineering and Computer Science, KTH Royal Institute of Technology, Stockholm, Sweden\\
\IEEEauthorrefmark{3}Mobile Communications and Computing, RWTH Aachen University, Germany
\\
Email:
\{milad.ganjalizadeh, hossein.shokri.ghadikolaei, johan.haraldson\}@ericsson.com,
petrovam@kth.se}
%\\
%\and
\vspace{-1cm}
%\Mark{$\ddagger$} \IEEEauthorblockA{University of California Davis\\
%Email: xbwang@ucdavis.edu }
}
\markboth{}%
{Shell \MakeLowercase{\textit{et al.}}: Bare Demo of IEEEtran.cls for IEEE Communications Society Journals}

\maketitle
	
\begin{abstract}
Distributed artificial intelligence (AI) has recently accomplished tremendous breakthroughs in various communication services, ranging from fault-tolerant factory automation to smart cities. When distributed learning is run over a set of wireless connected devices, random channel fluctuations, and the incumbent services simultaneously running on the same network affect the performance of distributed learning.
In this paper, we investigate the interplay between distributed AI workflow and ultra-reliable low latency communication (URLLC) services running concurrently over a network. 
Using 3GPP compliant simulations in a factory automation use case, we show the impact of various distributed AI settings (e.g., model size and the number of participating devices) on the convergence time of distributed AI and the application layer performance of URLLC. Unless we leverage the existing 5G-NR quality of service handling mechanisms to separate the traffic from the two services, our simulation results show that the impact of distributed AI on the availability of the URLLC devices is significant. Moreover, with proper setting of distributed AI (e.g., proper user selection), we can substantially reduce network resource utilization, leading to lower latency for distributed AI and higher availability for the URLLC users. Our results provide important insights for future 6G and AI standardization.
\end{abstract}
	
% Note that keywords are not normally used for peerreview papers.
\IEEEpeerreviewmaketitle

%%%%%%%%%%%%%%%%%%%%%
%%%% New Section %%%%
%%%%%%%%%%%%%%%%%%%%%
\section{Introduction}\label{sec:intro}
Future 6G networks are envisioned as an unprecedented evolution from connected things to connected intelligence, thereby serving as the backbone of a cyber-physical world with the integration of connected devices, intelligence, and humans \cite{surveyEdgeShi}. Numerous 6G \gls{ai} applications have emerged recently to improve efficiency and system performance in many vertical sectors, such as industrial automation \cite{ganjalizadehGcOrch}, autonomous driving \cite{li2019federated}, and enhanced mobile broadband \cite{cascadedFL}. Centralized training of the models can be impractical in many wireless communication applications because of (i) the distributed nature of the data generated/collected by mobile devices, (ii) privacy concerns on sharing the local data with a central server, especially when the computational server is managed by a third party operator, and (iii) limited wireless resources (in terms of bandwidth and power). Therefore, privacy-preserving distributed \gls{ai} techniques have become the cornerstone of recent advancements in \gls{ai} applications over wireless networks. In most distributed training algorithms, a set of devices upload their local updates (in terms of e.g., gradients in distributed stochastic gradient descent \cite{duttaKsync}, or local models in \gls{fl} \cite{fl}) via \gls{ul} channel to a master node (or a set of nodes) that maintains global parameters. Once the master node updates the global model, it shares them with the devices in a \gls{dl} channel.

Nevertheless, the \gls{ul}/\gls{dl} transmissions of \gls{ai} gradients/models are prone to errors and delays by the wireless channel, impacting the learning performance in terms of convergence time  \cite{saadConvergence}. There can be two remarks for distributed training on wireless systems. On the one hand, in addition to the number of iterations, the convergence time of the distributed training algorithm depends on the amount of time in which global model parameters are transmitted to the devices, trained locally, and transmitted back to the master node. On the other hand, the general perception is that increasing the \gls{ai} model size improves the training accuracy \cite{mobileNets}, given enough data samples and a proper training approach that reduces over-fitting. However, using a larger \gls{ai} model means longer communication and computation time, resulting in higher convergence time \cite{saadConvergence}. Higher AI communication overheads may also be detrimental for other communication services running in parallel to the AI. The tighter the requirements of the underlying service, the harder to design smooth coexistence.

\Gls{urllc} is characterized by strict requirements in terms of latency, which could be as short as $500$\,$\mu$s, and availability, which could be as high as $99.9999999$ \cite{3GPP22104}. Regarded as the most challenging use case in \gls{5g} and beyond 5G, this type of service is supposed to enable challenging applications (e.g., factory automation or autonomous intelligent transport systems \cite{urllcApp}) that have not been feasible in preceding generations of wireless communications.
% We can have two observations here:
% (i) Due to the statistical correlation of the data among different devices, having updates from every device may not be needed, and the missing parameters/gradients of some devices can be approximated by the ones of other devices. This suggests the possibility of user selection, as done, for instance, in federated learning.  
% (ii) Due to the local smoothness of most objective functions used for the training, the training algorithm (and therefore the master node) may tolerate getting late updates from some devices or even use some outdated parameters/gradients to update global parameters in the master node. Clearly, the delay between the current iteration number and the last time the orchestrator received parameters/gradients of every device should be bounded to guarantee convergence of the algorithm. 

As distinct services, the performance of both \gls{urllc} and distributed \gls{ai} over wireless networks have been widely investigated in the existing literature. However, the coexistence of \gls{urllc}, with its stringent requirements, and distributed \gls{ai} workflow, with its unique traffic model and performance characteristics, have not yet been discussed in the literature. Such coexistence introduces new fundamental challenges as well as unique trade-offs between \gls{urllc} latency and availability on the one hand, and convergence time and accuracy of distributed \gls{ai} on the other hand. Notice that there are fundamental differences between AI service and other traditional communication traffic. In particular: 
\begin{itemize} 
\item Due to the statistical correlation of the data among different nodes, the missing gradients/models of some nodes can be approximated by the ones of other nodes, and we can run distributed learning workflow seamlessly. 
\item Due to the local smoothness of most objective functions used for the training, the training algorithm (and therefore the master node) may tolerate getting late updates from some devices or even use some outdated local gradients/models to update global parameters \cite{agarwal2011distributed}. 
\end{itemize}
These are unique characteristics of distributed learning that distinguishes it from other traditional communication services. 

In this paper, we address the following research questions:
\begin{itemize}
   \item What are the trade-offs between \gls{urllc} availability and distributed \gls{ai} convergence time and accuracy? 
    \item  How does the \gls{ai} setting (e.g., model size, the number of participating AI devices, or the number of extra workers providing resilience against straggling) impact its own and the \gls{urllc} performance? 

\end{itemize}
We address the above research questions with \gls{3gpp} compliant \gls{5g}-NR simulations using network and link level simulations. Within our simulations, we observe that the number of participating \gls{ai} devices can have a significant impact on \gls{urllc} performance. Moreover, we show that introducing a soft synchronous protocol, in which the master node initiates the global model update upon receiving local updates from a portion of the participating devices, can help reduce the training delay in the order of seconds. Our results are also an important contribution to the ongoing standardization activities on \gls{5g} and beyond 5G to support distributed \gls{ai} model distribution and transfer.

% \Hossein{please add some details on insights we get, like... We have observed that... We show that... It is important to consider .... Our results make important contribution to the ongoing standardization activity of....}

The rest of this paper is organized as follows. In Section~\ref{sec:PR}, we discuss related works and introduce \gls{urllc} and distributed \gls{ai} \glspl{kpi}. We describe our link level and network level simulations in Section~\ref{sec:simulation}. Section~\ref{sec:preformance} presents the main results and insights on the coexistence scenario. Section \ref{sec:conculsions} concludes the paper.

\textit{Notations:} 
Normal font $x$ or $X$, bold font $\bm{x}$ or $\bm{X}$, and uppercase calligraphic font $\mathcal{X}$ denote scalar, vector and set, respectively. We denote by $|\calX|$ the cardinality of set $\calX$, by $\mathds{1}\{x\}$ the indicator function taking 1 only when condition $x$ holds, and by $[N]$ set $\{1,2,\ldots,N\}$. 
%%%%%%%%%%%%%%%%%%%%%
%%%% New Section %%%%
%%%%%%%%%%%%%%%%%%%%%
\section{Background and System Model}\label{sec:PR}
\subsection{Distributed Learning}
Consider the problem of minimizing a sum of functions $\{f_{i}: \R^d \mapsto \R\}_{i \in [N]}$, with corresponding gradients $\{\nabla f_{i}: \R^d \mapsto \R^d\}_{i \in [N]}$:
\begin{equation}\label{eq: MainOptimProblem}
\bw^\star {\coloneqq} \min_{\bw\in\R^d} f(\bw) {=} \min_{\bw\in\R^d} \frac{1}{N}\sum\nolimits_{i \in [N]}f_i(\bw) \:.
\end{equation}
Such problems frequently arise in distributed learning where each $f_i$ could represent a local model. In practice, to parallel the computations or to preserve the privacy of local datasets, we use distributed algorithms to solve~\eqref{eq: MainOptimProblem}~\cite{Bottou2018SIAM}. That is, at iteration $k$, a subset of the workers compute and upload their gradients $\{\nabla f_i(\bw_k)\}_i$ to a master node, which updates the model and broadcasts the updated model parameters $\bw_{k+1}$ to the workers. Federated learning is another popular method in which the workers will run one or several local training passes and upload their local models afterward. The master node will then take a global average over them. The communication overhead is almost the same as uploading gradients \cite{li2019federated}. However, most of these uplink messages (gradients or local models) are redundant, carrying almost no additional information since they can be retrieved from their past communicated messages as well as messages of other devices \cite{ghadikolaei2021lena}. Forcing some of them to remain silent would 1) reduce uplink interference to other users, 2) increase throughput, and 3) improve latency.

%In a conventional FL algorithm \cite{li2019federated}, a random subset of workers, say $n$ out of all $N$ clients, will be selected at every iteration and their messages will be used to update the global model. 
%Reference \cite{ji2020dynamic} proposed an algorithm to adjust $n$ at every iteration.
%, say $n_k$ at iteration $k$. Still, they randomly pick those set of $n_k$ clients. 
%References~\cite{mcmahan2017communication,karimireddy2020scaffold,ghadikolaei2021lena,chen2018lag} proposed various approaches to eliminate some unnecessary uploads. 
%In particular, every device locally tracks the changes of its gradients and uploads the new gradients/parameters only if the norm-2 of the changes is large enough. 
%However, none of those works study or optimize the interplay between distributed learning and other parallel communication services.    
In conventional synchronous distributed training methods, the master node should wait to receive the local updates from all participating devices, leading to a considerable inoperative time in the master node waiting for stragglers. To tackle the straggler's problem, in $n$-synch approaches, the master node only waits for a subset of participating devices, say $n$ out of all $N$ devices, and updates the global model using their messages at every iteration \cite{duttaKsync}. Nevertheless, vanilla $n$-synch-based methods add extra load on the underlying communication system, as they will ask all the devices to upload their data, and the master node starts its update with the first $n$ received data. Reference \cite{ji2020dynamic} proposed an algorithm to adjust $n$ at every iteration. References~\cite{fl,ghadikolaei2021lena,chen2018lag} proposed various approaches to eliminate some unnecessary uploads. However, none of those works study or optimize the interplay between distributed learning and other parallel communication services.
%In a distributed computation setting such as in \gls{iot} or edge learning, running each iteration of the algorithm involves some costs $c_i$, which could correspond to the number of bits (or energy or latency) needed to transmit $\{\nabla f_i(\bw_k)\}_i$ or the computational resources needed to compute $\{\nabla f_i(\bw_k)\}_i$. These costs become of paramount importance when we implement \gls{ml} and distributed optimization algorithms on bandwidth-limited  networks. The problem gets more challenging in wireless networks with additional limitations on communication range, battery, interference, and low computational capabilities. In such networks, tight requirements on low end-to-end latency (in autonomous driving), low energy usage (\gls{iot}), and high reliability (remote industrial operation) may render the ultimate solution, and consequently the distributed algorithm, useless~\cite{jeschke2017industrial,Jiang2019LowLatency}\Milad{@hossein: what was supposed to be here? Besides, below is the original plan, what do you think is missing?

% \Milad{Do you think we should add more in
% \begin{itemize}
%     \item convergence rate of \gls{sgd} and \gls{fl}
%     \item communication efficiency
% \end{itemize}
% }
\subsection{System Model}\label{sec:systemModel}
We consider an industrial automation scenario where a set of $\mathcal{M}{\coloneqq}[M]$ industrial devices in the factory hall execute different functions that enable automated production. The communication system should timely and reliably deliver (i) monitoring data to \glspl{gnb} and (ii) computed or emergency control commands to the actuators. However, we consider application-layer performance for \gls{urllc} service, implying that consecutive failures that are shorter than survival time ($T_{\mathrm{sv}}$) do not affect the end-to-end performance. %Besides, we assume there is one \gls{gnb}, configured with a three-sector cell setting. 

For simplicity, we assume that \gls{ai} devices are distinct from industrial devices, and there exist a set of participating \gls{ai} devices $\mathcal{U} {\coloneqq} [N]$ serving a background \gls{ai} task. Moreover, we assume that the \gls{ai} master node requires to receive the relevant local information from $n$ out of these $N$ participating \gls{ai} devices to update its global model. For simplicity, we define $\eta{\coloneqq}\frac{n}{N}$. Therefore, we obtain fully synchronous and $n$-synch distributed training for $\eta{=}1$ and $\eta{<}1$, respectively.

%\subsection{Resource Management}
To manage the coexistence of two services,  
%the fifth generation of wireless networks (\gls{5g}) and beyond \gls{5g} are envisioned to support a mixture of heterogeneous use cases (e.g., industrial automation and smart agriculture), which impose a broad range of performance requirements. In the coexistence scenarios, as here, 
where the priority of services are inherently different, \gls{5g} and beyond 5G envision two approaches. The first approach, employed in this paper, is to use the existing standardized protocols in 5G-NR for \gls{qos} handling. In this case, each connected device is assigned with one or several \gls{qos} flows and data radio bearers, where the former is set in the core network, depending on the service \gls{qos} requirements. For example, in our scenario, the traffic from/to \gls{urllc} devices is set to have high priority \gls{qos} flow to ensure low latency, whilst the traffic from/to \gls{ai} devices is set to have low priority \gls{qos} flow. Each (or several) of these \gls{qos} flows are then mapped to a data radio bearer in the radio access network. In \gls{gnb} and devices, there is an associated \gls{rlc} buffer to each data radio bearer, and in our case, with strict priority scheduling \cite{dahlman5GNr}. The second approach, as we foresee, is to have separate slices for \gls{urllc} and distributed \gls{ai}, resulting in full resource separation (e.g., in terms of bandwidth).

%\subsection{Distributed Algorithm}\label{sec:distAlg}
To model a distributed learning task, we consider a network of $N$ \gls{ai} devices that cooperatively solve a distributed learning problem. Iteration $k$ of an abstract distributed algorithm reads:
\vspace{-2mm}
\begin{subequations}\label{eq:dAI}
\begin{align}
    &\bw_{k+1}{=}A\left(\bc_{i,k+1}, \bw_k\right),\quad \mbox{for} \quad \forall i \in \calU_n \label{eq:globalUp}\\
    & \bc_{i,k+1}{=}C_i\left(\bw_{k+1}\right),\quad \mbox{for} \quad \forall i \in \calU \label{eq:localUp}
\end{align}
\end{subequations}
where function $A$ represents an algorithm update of the decision variable $\bw_k$, function $C_i$ picks out the relevant information, $\bc_{i,k}$, that node $i$ uploads to the server to run the algorithm. This general algorithmic framework covers many \gls{ml} algorithms, including federated learning and distributed stochastic gradient descent, with or without data compression. For example, when $C_i$ returns a stochastic gradient, say $\hat{\nabla} f_i(\bw_{k+1})$, and $A=\bw_k - \alpha \sum_i \hat{\nabla} f_i(\bw_{k+1})/n$ for some positive step size $\alpha$, we recover $n$-synch and synchronous distributed stochastic gradient descent for $n({<} N)$ and $n({=} N)$. When $C_i$ returns an updated local model parameters of \gls{ai} device $i$ and $A$ takes an averaging step over a subset of $n ({\leq} N)$ \gls{ai} devices, we recover FL ($n$-synch or synchronous).

In the next section, we use these models to formulate performance metrics.
\section{Performance Metrics}\label{sec:PM}
\subsection{\gls{urllc}: Application Layer Availability}\label{sec:urllc}
The reliability of \gls{urllc} service is typically identified by application layer availability, where such \gls{kpi} is measured at the application layer of the end device \cite{3GPP22104}. In terms of reliability, the main difference between the observed performance on the application layer with the observed performance on the network layer is driven by a system parameter called survival time, $T_{\mathrm{sv}}$. Survival time is a duration of time for which the application layer can tolerate failures in the communication system without any performance degradation in availability \cite{ganjalizadehPimrcTranslating}. Let us denote the communication system state variable by a Bernoulli state variable $X_i{\left(t\right)}$, where $X_i{\left(t\right)}$ for the $i$th \gls{urllc} device is zero if the last packet reception at network layer has failed, either because it could not be decoded at the lower layers or the packet has been received after its corresponding delay bound.
Consequently, we define the per-device application layer state variable, $Y{\left(t\right)}$ as 
\begin{equation}\label{eq:appState}
Y_i{\left(t\right)}{\coloneqq}
    \begin{cases}
      0 & \mathrm{if} \int_{\tau=t-T_{\mathrm{sv}}}^t X_i{\left(\tau\right)}d\tau {=} 0,\\
      1 &  \mathrm{otherwise}. \\
    \end{cases}
\end{equation}
Therefore, we can define the long-term availability of the $i$th industrial device as \cite{ganjalizadehGcOrch}
\begin{equation}
    a_i{\coloneqq} \lim_{t\to \infty}\Pr\left\{Y_i\left(t\right) {=} 1\right\} {=} \lim_{T\to\infty}\frac{1}{T}\int_{t=0}^T Y_i(t) dt.
\end{equation}
The availability can be estimated over a short time using
\begin{equation}
    \Bar{a}_i{\coloneqq} \frac{1}{T}\int_{t=0}^T Y_i(t) dt. \label{eq:availSim}
\end{equation}
In \gls{urllc} applications, the requirement is often defined in the form of \cite{popovskiUrllc}
\begin{equation}\label{eq:availReq}
    \Pr\left\{a_i \leq a_i^{\mathrm{req}}\right\}\leq\gamma, \forall i\in\mathcal{M},\
\end{equation}
where $a_i^{\mathrm{req}}$ is the availability requirement for \gls{urllc} device $i$, and $\gamma$ is the sensitivity of the application to $a_i^{\mathrm{req}}$.
%where $a_i^{\mathrm{req}}$ and $\gamma_i$ are the availability requirement and the sensitivity to availability requirement for \gls{urllc} device $i$, respectively.
\subsection{Distributed AI: Training Delay}\label{sec:urllc}
The convergence time of the distributed \gls{ai} is bounded by the communication and processing latency \cite{saadConvergence}. Let us assume $\mathcal{U}_n \subseteq \mathcal{U}$ where $|\mathcal{U}_n| {=} n$. Based on the abstract distributed algorithm in Section \ref{sec:systemModel}, the \gls{ai} training delay in the master node for $k$th iteration, $d_k^{\mathrm{AI}}$, can be derived as
\begin{equation}\label{eq:aiDelay}
d_k^{\mathrm{AI}} = \min_{\forall\mathcal{U}_n\subseteq \mathcal{U} }\left[\max_{\forall i\in\mathcal{U}_n}\left(d_{i,k}^{\mathrm{D}}+d_{i,k}^{\mathrm{pr}}+d_{i,k}^{\mathrm{U}}+d_k^{\mathrm{pr}}\right)\right],
\end{equation}
where $d_{i,k}^{\mathrm{D}}$, $d_{i,k}^{\mathrm{pr}}$, and $d_{i,k}^{\mathrm{U}}$, are the latency of \gls{dl} global model broadcasting, local training (represented in \eqref{eq:localUp}), and \gls{ul} transmission of local gradients/models for $k$th iteration of $i$th device, respectively. It is worth noting that $d_{i,k}^{\mathrm{D}}$ and $d_{i,k}^{\mathrm{U}}$ include the  transmission processing, payload transmission, and queuing delay, which is determined by the number of devices sharing the same time-frequency resources. Besides, $d_k^{\mathrm{pr}}$ is the $k$th iteration processing delay required to perform the global model update on the master node, represented in \eqref{eq:globalUp}. Thus, in equation \eqref{eq:aiDelay}, for each subset, the maximum aggregated communication and processing delay is calculated among devices. Then, among subsets, $d_k^{\mathrm{AI}}$ is determined by picking the subset with the lowest delay.
%in equation \eqref{eq:aiDelay}, 
%The $d_k^{\mathrm{AI}}$ is then from the subset that, among all the subsets with cardinality of $n$, has the minimum of the maximum aggregated communication and processing delay among its corresponding users for $k$th iteration.

%%%%%%%%%%%%%%%%%%%%%
%%%% New Section %%%%
%%%%%%%%%%%%%%%%%%%%%
\section{Simulation Methodology}\label{sec:simulation}
\begin{figure}[t]
	\centering
	\includegraphics[width=.75\columnwidth,keepaspectratio]{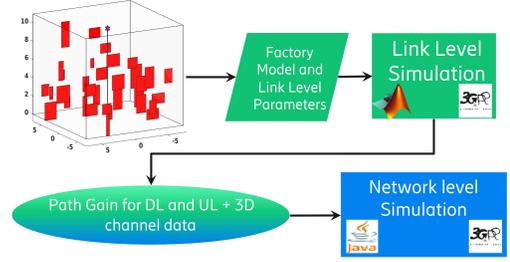}
	\caption{The simulation setup.}
	\label{fig:factory}
	\vspace{-6mm}
\end{figure}
For simulating the \gls{urllc} and distributed \gls{ai} coexistence deployment, we performed both link level and network level simulations for a factory automation scenario (as shown in \figurename\,\ref{fig:factory}). More explicitly, we designed a 3D model of a small factory of size $15\times15\times11$\,m$^3$ with a \gls{gnb} in the middle at the height of $10$\,m. We assumed that the \gls{gnb} is configured with a 3-sector cell setting. In link level simulations, we modeled blockers with the width and height that are uniformly selected from the range of $[0.5, 2]$\,m and $[1, 3]$\,m, respectively, and positioned them randomly inside the factory.
%walls We distributed the blockers randomly such that the centers of $80$\% are positioned with the height between $[0.5, 5]$\,m and the rest are between $[5, 9]$\,m. 
We leveraged the blockage model B from \cite{3GPP38901} to determine the multipath attenuation caused by each of the blockers using a knife-edge diffraction method. The path gain matrix and 3D channel data for all possible devices' locations are then imported to a network level simulator in which we simulated \gls{phy}, \gls{mac}, and above layers in a multi-cell multi-user scenario. 

The network level simulator is event-based, \gls{3gpp} compliant, and operates at \gls{ofdm} symbol resolution. We considered numerology one from \cite{dahlman5GNr}, implying that each slot and symbol are $0.5$\,ms and $33.33$\,$\mu$s long, respectively. At each seed, the location of both \gls{urllc} and \gls{ai} devices are selected randomly. 
%However, the positions were kept fixed within one seed, and instead, to simulate the mobility, we assumed that the surrounding environment moves with a speed of $30$\,km/h. 
To ensure seamless training of distributed \gls{ai} until the end of a simulation, we considered \gls{rlc} in \gls{am} for distributed \gls{ai}. Nevertheless, the \gls{rlc} retransmissions are slow and unlikely to benefit \gls{urllc} packets with their tight delay bounds \cite{dahlman5GNr}. Accordingly, we configured the \gls{rlc} in \gls{um} for \gls{urllc} flow. Besides, we assumed strict priority scheduling where \gls{urllc} flow has higher priority than \gls{ai} flow, implying that \gls{ai} packets can not be scheduled unless there is no \gls{urllc} packet on the queues.

Upon transmission, one or several packets are drawn from the head of the corresponding \gls{rlc} buffer, depending on the selected modulation and coding scheme on lower layers. Alternatively, \gls{rlc} could perform segmentation of packets into smaller segments to fit them into transport blocks via which the packets are transmitted. Upon reception, the received instantaneous \gls{sinr} of each transport block (which depends on the radio channel and the dynamic interference of other devices' transmissions) determines an error probability. Consequently, the receptive \gls{rlc} entity reassembles successfully decoded segments and delivers them to the application layer. For availability calculation on the application layer, we considered a \gls{urllc} packet lost if it is not fully received before its corresponding delay bound, followed by applying $T_{sv}$ as in \eqref{eq:availSim} where $T$ is the duration of one simulation. \tablename\,\ref{tab:simSetup} presents the simulation parameters.

On the traffic modeling, we considered $10$ \gls{urllc} devices. The \gls{urllc} traffic is represented by periodic \gls{ul} and \gls{dl} traffic, with delay bound of $6$\,ms and $2$\,ms as well as size of $64$ bytes and $80$ bytes, respectively, both with period $5$\,ms.
Motivated by \cite{3GPP22874AiModel}, we assumed that the shared \gls{dnn} architecture (i.e., used on the devices and the master node) follows MobileNets \cite{mobileNets}, a class of efficient \gls{dnn} models based on a streamlined architecture for mobile and embedded vision applications. We considered $x$\,MobileNet-$224$ in \cite{mobileNets}, where $x{\in}\{0.25, 0.5, 0.75, 1\}$, implying that the \gls{dnn} model can have $0.5$, $1.3$, $2.6$, or $4.2$ million parameters, respectively.
To model the distributed \gls{ai} traffic, we assumed \gls{fl} and 32 bits quantization for each model parameter, implying that each model (local or global) can be represented by a size of $2$\,MB, $5.2$\,MB, $10.4$\,MB $16.8$\,MB in the case of $0.25$\,MobileNet, $0.5$\,MobileNet, $0.75$\,MobileNet, and $1$\,MobileNet, respectively.
\begin{table}[t]
\centering
\caption{Simulation Parameters}
\label{tab:simSetup}
\scalebox{0.7}{\hspace{0.5mm}\begin{tabular}{ | l|l|  }
	\hline
	\textbf{Parameter}& \textbf{Value}\\
	\hline
	Deployment   & $1$ \gls{gnb}, $3$ cells\\
	Duplex/Carrier frequency   & FDD/2600\,MHz\\
%	Blocker width   & $[0.5, 2]$\,m\\
%	Blocker height   & $[1, 3]$\,m\\
	Blocker's density   & $0.15$\,blocker/m$^2$\\
	\gls{gnb} antenna height&   $10$\,m \\
	Devices' height &$1.5$\,m \\
	Carrier frequency & $2.6$\,GHz \\
	Bandwidth&   $40$\,MHz\,\\
	TTI length/Subcarrier spacing& $0.5$\,ms/$30$\,KHz  \\
	UL/DL transmit power &   $0.2$\,W/$0.5$\,W  \\
	Max\,num\,of\,\gls{ul}/\gls{dl}\,\gls{urllc}\,Trans. (\gls{mac}) & $3/2$ \\
	Max\,num\,of\,\gls{ul}/\gls{dl}\,\gls{ai}\,Trans. (\gls{mac}) & $10/10$\\
	Max\,num\,of\,\gls{ul}/\gls{dl}\,\gls{ai}\,Trans. (\gls{rlc}) & $8/8$ \\
	\gls{ul}/\gls{dl}\,\gls{urllc}\,delay bound& $6/2$\,ms \\
	\gls{ul}/\gls{dl} \gls{urllc} Survival time ($T_{\mathrm{sv}}$)& $5/5$\,ms \\
	Simulation time &  $100$\,s\\
	\hline
\end{tabular}}
 \vspace{-4.4mm}
\end{table}
% \Milad{
% \begin{itemize}
%     \item Modulation from 3GPP TS 38.214 Table 5.1.3.1.3
%     \item check if .mat can be shared
% \end{itemize}
% }
%%%%%%%%%%%%%%%%%%%%%
% NEW SECTION
%%%%%%%%%%%%%%%%%%%%%
\vspace{-3mm}
\section{Results and Discussion}\label{sec:preformance}
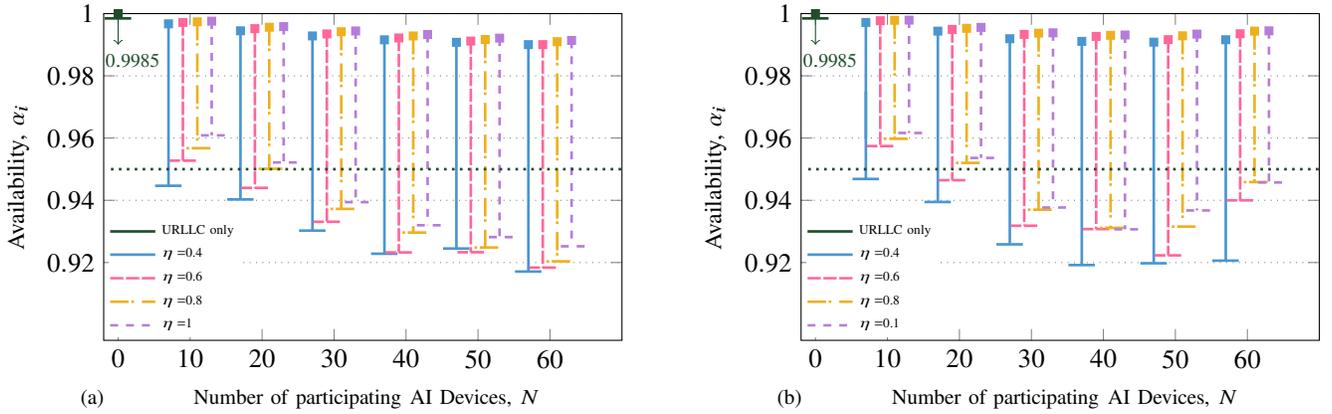
\begin{figure*}[t]
    \hspace{0mm}
    \begin{subfigure}[t]{0.47\textwidth}
	    \centering
    	% This file was created by matlab2tikz.
%
%The latest updates can be retrieved from
%  http://www.mathworks.com/matlabcentral/fileexchange/22022-matlab2tikz-matlab2tikz
%where you can also make suggestions and rate matlab2tikz.
%
\pgfplotsset{major grid style = {dotted, gray}}
\pgfplotsset{minor grid style = {dotted, gray}}

\begin{tikzpicture}

\begin{axis}[%
width=0.45*6.028*1in,
height=0.4*4.754*0.9in,
at={(0.753in,0.47in)},
scale only axis,
xmin=-2,
xmax=70,
xtick={0,10,20,30,40,50,60},
ytick={0.92,0.94,0.96,0.98,1},
ylabel={\small Availability, $\alpha_{i}$},
ymin=0.895,
ymax=1,
% xmajorgrids,
ymajorgrids,
legend style={at={(0.01,0.01)}, anchor=south west, legend cell align=left, align=left, draw=none,fill=white, inner sep=0.001pt, outer sep=0.001pt, row sep=0pt}
]
ylabel style={font=\color{white!15!black}},
ylabel={\small Availability mean/5th perc.},
axis background/.style={fill=white}
]
\addplot [color=mycolor1, only marks, mark size=1.5pt, mark=square*, mark options={solid, mycolor1}, forget plot]
 plot [error bars/.cd, y dir=minus, y explicit, error bar style={line width=1.0pt}, error mark options={line width=1.0pt, mark size=5.0pt, rotate=90}]
 table[row sep=crcr, y error plus index=2, y error minus index=3]{%
7	0.996709870388834	0	0.0520438683948156\\
17	0.994466600199402	0	0.0541874376869391\\
27	0.992821535393819	0	0.0625623130608176\\
37  0.991575274177468	0	0.0687437686939182\\
47  0.990777666999003	0	0.0662761714855433\\
57  0.990029910269192	0	0.0729312063808575\\
};

\addplot [color=mycolor2, only marks, mark size=1.5pt, mark=square*, mark options={solid, mycolor2}, forget plot]
 plot [error bars/.cd, y dir=minus, y explicit, error bar style={dash pattern={on 5pt off 1pt}, line width=1.0pt}, error mark options={line width=1.0pt, mark size=5.0pt, rotate=90}]
 table[row sep=crcr, y error plus index=2, y error minus index=3]{%
9	0.99715852442672	0	0.0444167497507477\\
19  0.995164506480558	0	0.0511714855433698\\
29  0.993469591226321	0	0.0603688933200399\\
39  0.992173479561316	0	0.0689182452642074\\
49  0.991176470588235	0	0.0678713858424727\\
59  0.990029910269192	0	0.0716600199401795\\
};
\addplot [color=mycolor3, only marks, mark size=1.5pt, mark=square*, mark options={solid, mycolor3}, forget plot]
 plot [error bars/.cd, y dir=minus, y explicit, error bar style={dash pattern={on 7pt off 1pt on 1pt off 3pt}, line width=1.0pt}, error mark options={line width=1.0pt, mark size=5.0pt, rotate=90}]
 table[row sep=crcr, y error plus index=2, y error minus index=3]{%
11	0.997357926221336	0	0.0406031904287139\\
21  0.995613160518445	0	0.0454885343968094\\
31  0.994167497507478	0	0.0569292123629113\\
41  0.992821535393819	0	0.063185443668993\\
51  0.991674975074776	0	0.0668494516450647\\
61  0.990977068793619	0	0.0705882352941176\\
};
\addplot [color=mycolor4, only marks, mark size=1.5pt, mark=square*, mark options={solid, mycolor4}, forget plot]
 plot [error bars/.cd, y dir=minus, y explicit, error bar style={dash pattern={on 3pt off 3pt}, line width=1.0pt}, error mark options={line width=1.0pt, mark size=5.0pt, rotate=90}]
 table[row sep=crcr, y error plus index=2, y error minus index=3]{%
13	0.997557328015952	0	0.0366899302093718\\
23  0.995812562313061	0	0.0436440677966102\\
33  0.994416749750748	0	0.0550099700897309\\
43  0.993270189431705	0	0.0612911266201396\\
53  0.992123629112662	0	0.0639581256231306\\
63  0.991375872382851	0	0.0661515453639083\\
};

\addplot [color=mycolor5, only marks, mark size=1.5pt, mark=square*, mark options={solid, mycolor5}, forget plot]
 plot [error bars/.cd, y dir=minus, y explicit, error bar style={line width=1.0pt}, error mark options={line width=1.0pt, mark size=5.0pt, rotate=90}]
 table[row sep=crcr, y error plus index=2, y error minus index=3]{%
0	1	0	0.0015453639082752\\
};
\draw [->,mycolor5] (0,0.9985) -- (0,0.99) ;
\node[text width=1mm,fill=none, mycolor5] at (-1.2,0.985) {\scriptsize $0.9985$};

 \addplot[const plot, color=mycolor5,mark repeat=25, line width=1.0pt, mark size=1.0pt, mark=none, mark options={solid, fill=mycolor5, mycolor5}] table[row sep=crcr] {%
0	0.99999\\
0	0.99995\\
0	0.99991\\
0	0.9999\\
};
\addlegendentry{\tiny URLLC only}

\addplot[const plot, color=mycolor1,mark repeat=25, line width=1.0pt, mark size=1.0pt, mark=none, mark options={solid, fill=mycolor1, mycolor1}] table[row sep=crcr] {%
7	0.996709870388834\\
7	0.9967\\
7	0.9966\\
7	0.9965\\
};
\addlegendentry{\tiny $\eta=$0.4}

\addplot[dash pattern={on 5pt off 1pt}, color=mycolor2,mark repeat=25, line width=0.95pt, mark size=1.0pt, mark=none, mark options={solid, fill=mycolor2, mycolor2}] table[row sep=crcr] {%
9	0.99715852442672\\
9	0.99715\\
9	0.99713\\
9	0.99711\\
};
\addlegendentry{\tiny $\eta=$0.6}

\addplot[dash pattern={on 7pt off 1pt on 1pt off 3pt}, color=mycolor3,mark repeat=25, line width=0.95pt, mark size=1.0pt, mark=none, mark options={solid, fill=mycolor3, mycolor3}] table[row sep=crcr] {%
11	0.997357926221336\\
11	0.9973\\
11	0.9974\\
11	0.9973\\
};
\addlegendentry{\tiny $\eta=$0.8}

\addplot[dash pattern={on 3pt off 3pt}, color=mycolor4,mark repeat=25, line width=0.95pt, mark size=1.0pt, mark=none, mark options={solid, fill=mycolor4, mycolor4}] table[row sep=crcr] {%
13	0.997557328015952\\
13	0.99755\\
13	0.99754\\
13	0.99753\\
};
\addlegendentry{\tiny $\eta=$1}

 \addplot[dotted, color=vColor, line width=1.0pt] table[row sep=crcr] {%
 -1	0.95\\
 0	0.95\\
 10	0.95\\
 20	0.95\\
 30	0.95\\
 40	0.95\\
 50	0.95\\
 60	0.95\\
 70	0.95\\
 };
\end{axis}
\end{tikzpicture}%
    	\vspace{-5mm}\hspace{7mm}
    	\caption{\hspace{10mm} Number of participating AI Devices, $N$}
        \label{fig:avail025}
    \end{subfigure}
    \hspace{5mm}
    \begin{subfigure}[t]{0.47\textwidth}
	\centering
    	% This file was created by matlab2tikz.
%
%The latest updates can be retrieved from
%  http://www.mathworks.com/matlabcentral/fileexchange/22022-matlab2tikz-matlab2tikz
%where you can also make suggestions and rate matlab2tikz.
%
\pgfplotsset{major grid style = {dotted, gray}}
\pgfplotsset{minor grid style = {dotted, gray}}

\begin{tikzpicture}

\begin{axis}[%
width=0.45*6.028*1in,
height=0.4*4.754*0.9in,
at={(0.753in,0.47in)},
scale only axis,
xmin=-2,
xmax=70,
xtick={0,10,20,30,40,50,60},
ytick={0.92,0.94,0.96,0.98,1},
ylabel={\small Availability, $\alpha_{i}$},
ymin=0.895,
ymax=1,
% xmajorgrids,
ymajorgrids,
legend style={at={(0.01,0.01)}, anchor=south west, legend cell align=left, align=left, draw=none,fill=white, inner sep=0.001pt, outer sep=0.001pt, row sep=0pt}
]
ylabel style={font=\color{white!15!black}},
ylabel={\small Availability mean/5th perc.},
axis background/.style={fill=white}
]
\addplot [color=mycolor1, only marks, mark size=1.5pt, mark=square*, mark options={solid, mycolor1}, forget plot]
 plot [error bars/.cd, y dir=minus, y explicit, error bar style={line width=1.0pt}, error mark options={line width=1.0pt, mark size=5.0pt, rotate=90}]
 table[row sep=crcr, y error plus index=2, y error minus index=3]{%
7	0.99715852442672	0	0.0502991026919242\\
17	0.994366899302094	0	0.0549102691924228\\
27	0.991924227318046	0	0.0660767696909272\\
37  0.991076769690927	0	0.0718843469591225\\
47  0.990827517447657	0	0.0710618145563311\\
57  0.991600199401795	0	0.07098703888335\\
};

\addplot [color=mycolor2, only marks, mark size=1.5pt, mark=square*, mark options={solid, mycolor2}, forget plot]
 plot [error bars/.cd, y dir=minus, y explicit, error bar style={dash pattern={on 5pt off 1pt}, line width=1.0pt}, error mark options={line width=1.0pt, mark size=5.0pt, rotate=90}]
 table[row sep=crcr, y error plus index=2, y error minus index=3]{%
9	0.997706879361914	0	0.0402791625124627\\
19  0.994915254237288	0	0.0484546360917248\\
29  0.993320039880359	0	0.0614905284147557\\
39  0.992671984047856	0	0.061914257228315\\
49  0.991625124626122	0	0.0693170488534396\\
59  0.993519441674975	0	0.0535393818544367\\
};
\addplot [color=mycolor3, only marks, mark size=1.5pt, mark=square*, mark options={solid, mycolor3}, forget plot]
 plot [error bars/.cd, y dir=minus, y explicit, error bar style={dash pattern={on 7pt off 1pt on 1pt off 3pt}, line width=1.0pt}, error mark options={line width=1.0pt, mark size=5.0pt, rotate=90}]
 table[row sep=crcr, y error plus index=2, y error minus index=3]{%
11	0.997856430707876	0	0.038085742771685\\
21  0.995264207377866	0	0.0432452642073778\\
31  0.993718843469591	0	0.0567298105682952\\
41  0.993020937188435	0	0.061864406779661\\
51  0.992871385842473	0	0.0613160518444665\\
61  0.994366899302094	0	0.048504486540379\\
};
\addplot [color=mycolor4, only marks, mark size=1.5pt, mark=square*, mark options={solid, mycolor4}, forget plot]
 plot [error bars/.cd, y dir=minus, y explicit, error bar style={dash pattern={on 3pt off 3pt}, line width=1.0pt}, error mark options={line width=1.0pt, mark size=5.0pt, rotate=90}]
 table[row sep=crcr, y error plus index=2, y error minus index=3]{%
13	0.99790628115653	0	0.0362662013958126\\
23  0.995563310069791	0	0.0419491525423729\\
33  0.993818544366899	0	0.0561316051844467\\
43  0.993120638085743	0	0.0624376869391825\\
53  0.993419740777667	0	0.056679960119641\\
63  0.994466600199402	0	0.048728813559322\\
};

\addplot [color=mycolor5, only marks, mark size=1.5pt, mark=square*, mark options={solid, mycolor5}, forget plot]
 plot [error bars/.cd, y dir=minus, y explicit, error bar style={line width=1.0pt}, error mark options={line width=1.0pt, mark size=5.0pt, rotate=90}]
 table[row sep=crcr, y error plus index=2, y error minus index=3]{%
0	1	0	0.0015453639082752\\
};
\draw [->,mycolor5] (0,0.9985) -- (0,0.99) ;
\node[text width=1mm,fill=none, mycolor5] at (-1.2,0.985) {\scriptsize $0.9985$};

 \addplot[const plot, color=mycolor5,mark repeat=25, line width=1.0pt, mark size=1.0pt, mark=none, mark options={solid, fill=mycolor5, mycolor5}] table[row sep=crcr] {%
0	0.99999\\
0	0.99995\\
0	0.99991\\
0	0.9999\\
};
\addlegendentry{\tiny URLLC only}

\addplot[const plot, color=mycolor1,mark repeat=25, line width=1.0pt, mark size=1.0pt, mark=none, mark options={solid, fill=mycolor1, mycolor1}] table[row sep=crcr] {%
7	0.96\\
7	0.965\\
7	0.97\\
7	0.975\\
};
\addlegendentry{\tiny $\eta=$0.4}

\addplot[dash pattern={on 5pt off 1pt}, color=mycolor2,mark repeat=25, line width=0.95pt, mark size=1.0pt, mark=none, mark options={solid, fill=mycolor2, mycolor2}] table[row sep=crcr] {%
9	0.997856430707876\\
9	0.99785\\
9	0.99783\\
9	0.99781\\
};
\addlegendentry{\tiny $\eta=$0.6}

\addplot[dash pattern={on 7pt off 1pt on 1pt off 3pt}, color=mycolor3,mark repeat=25, line width=0.95pt, mark size=1.0pt, mark=none, mark options={solid, fill=mycolor3, mycolor3}] table[row sep=crcr] {%
11	0.997706879361914\\
11	0.996\\
11	0.9955\\
11	0.9950\\
};
\addlegendentry{\tiny $\eta=$0.8}

\addplot[dash pattern={on 3pt off 3pt}, color=mycolor4,mark repeat=25, line width=0.95pt, mark size=1.0pt, mark=none, mark options={solid, fill=mycolor4, mycolor4}] table[row sep=crcr] {%
13	0.99790628115653\\
13	0.9979\\
13	0.9975\\
13	0.9973\\
};
\addlegendentry{\tiny $\eta=$0.1}

 \addplot[dotted, color=vColor, line width=1.0pt] table[row sep=crcr] {%
 -1	0.95\\
 0	0.95\\
 10	0.95\\
 20	0.95\\
 30	0.95\\
 40	0.95\\
 50	0.95\\
 60	0.95\\
 70	0.95\\
 };

\end{axis}
\end{tikzpicture}%
    	\vspace{-5mm}\hspace{7mm}
    	\caption{\hspace{10mm} Number of participating AI Devices, $N$ }
        \label{fig:avail1}
    \end{subfigure}
    \vspace{-1mm}
	\caption{Median\,(square) and $1$st percentile\,(line) availability of URLLC, when training (a)\,0.25\,MobileNet, and (b)\,1\,MobileNet, both with $32$\,bits quantization.}% for the \gls{dnn} model.}
	\label{fig:avail}
	\vspace{-4.5mm}
\end{figure*}
In this section, we evaluate the performance of \gls{urllc} and distributed \gls{ai} in our coexistence scenario. Besides evaluating the impact of $N$, we study $n$-synch on the coexistence scenario from two different perspectives:
\begin{itemize}
    \item \texttt{Eval1}: We compare cases with the same $N$, while changing $\eta$ from 0.4 to 1, to address the impact of tightening the requirements for a global update at the master node.  
    \item \texttt{Eval2}: We compare cases with the same $n$\!\! (${=}\eta\times N$), implying that the global model is aggregated from the same number of local models. Here, we evaluate the impact of $N{-}n$ extra devices available to reduce sensitivity to straggling devices.
    %\begin{greek}\texttt{h}\end{greek}\texttt{-N-Eval2}: $\eta\times N$ static: the global model is from same number of devices in the evaluation. $n$-synch, where the requirement is $n$ and $N-n$ extra devices to provide resilience to straggling. \texttt{1-20-Eval2}, \texttt{1-30-Eval2}, and \texttt{1-40-Eval2} to \texttt{0.4-50-Eval2}, \texttt{0.6-50-Eval2}, and \texttt{0.8-50-Eval2}, respectively.Assuming that $N-n$ devices are used as extra devices to leverage diversity
\end{itemize}
Besides, we use ${(N, \eta)}$ after a specific \texttt{Eval} to refer to $N$ and $\eta$ in a specific evaluation. For example, \texttt{Eval2}$\mathtt{\left(50{,}0.6\right)}$ refers to $N{=}50$ and $\eta{=}0.6$ for an \texttt{Eval2} experiment.

\figurename\,\ref{fig:avail} illustrates the impact of introducing distributed \gls{ai} workflow on \gls{urllc} service performance. \figurename\,\ref{fig:avail025} and \figurename\,\ref{fig:avail1} assume $0.25$\,MobileNet and $1$\,MobileNet \cite{mobileNets}, respectively. These figures show the median and the $1$st percentile availability of the \gls{urllc} devices, while having different number of \gls{ai} devices, $N$, and for different reception requirement, $\eta$. The dotted horizontal line represents $\alpha_i^{\mathrm{req}}$ in \eqref{eq:availReq}, where we assumed it is identical for all \gls{urllc} devices (i.e., $\alpha_i^{\mathrm{req}}{=}\alpha^{\mathrm{req}}{=}0.95$ for $\forall i\in \mathcal{M}$). We assumed $\gamma {=} 0.01$, hence, if the first percentile availability is higher than the dotted line, the availability requirement in \eqref{eq:availReq} is fulfilled. There is no notable difference observed between 0.25\,MobileNet  of \figurename\,\ref{fig:avail025} and 1\,MobileNet of \figurename\,\ref{fig:avail1}. As both figures confirm, introducing the \gls{ai} traffic causes a minimum of $0.037$ decrease in the $1$st percentile \gls{urllc} availability. This is shocking as  our deployment supports two nines availability in the \gls{urllc}-only scenario, whilst it barely fulfills availability of $0.96$ in the coexistence scenario. This significant availability reduction highlights the importance of intelligent user selection or slicing approaches to control the effect of AI traffic-induced interference on URLLC service. Moreover, \figurename\,\ref{fig:avail} shows that the availability of \gls{urllc} devices declines as we increase the number of \gls{ai} participants. Generally, a higher number of distributed \gls{ai} participants increases the interference, resulting in higher (i) packet error ratio and (ii) \gls{urllc} packet delays (since the scheduler requires to select lower modulation and coding scheme indices to cope with the higher interference). Consequently, the \gls{urllc} service does not meet its availability requirement, i.e., 0.95, when $N{\geq}30$. 

%Generally, the higher number of distributed \gls{ai} participants lead to higher interference, resulting in lower availability in \gls{urllc} devices. Such an escalation in interference increases (i) the decoding error probability and (ii) the \gls{urllc} packet delays (since the scheduler requires to select lower modulation and coding scheme indices to compensate for the target block error rate). 
% Such an increase in the training delay, given the same number of \gls{ai} devices, leads to lower interference and hence, higher availability for \gls{urllc} devices. 

\begin{figure*}[t]
    % \hspace{-1mm}
	\begin{subfigure}[b]{0.5\textwidth}
	\centering
    	\includegraphics[width=.99\columnwidth,keepaspectratio]{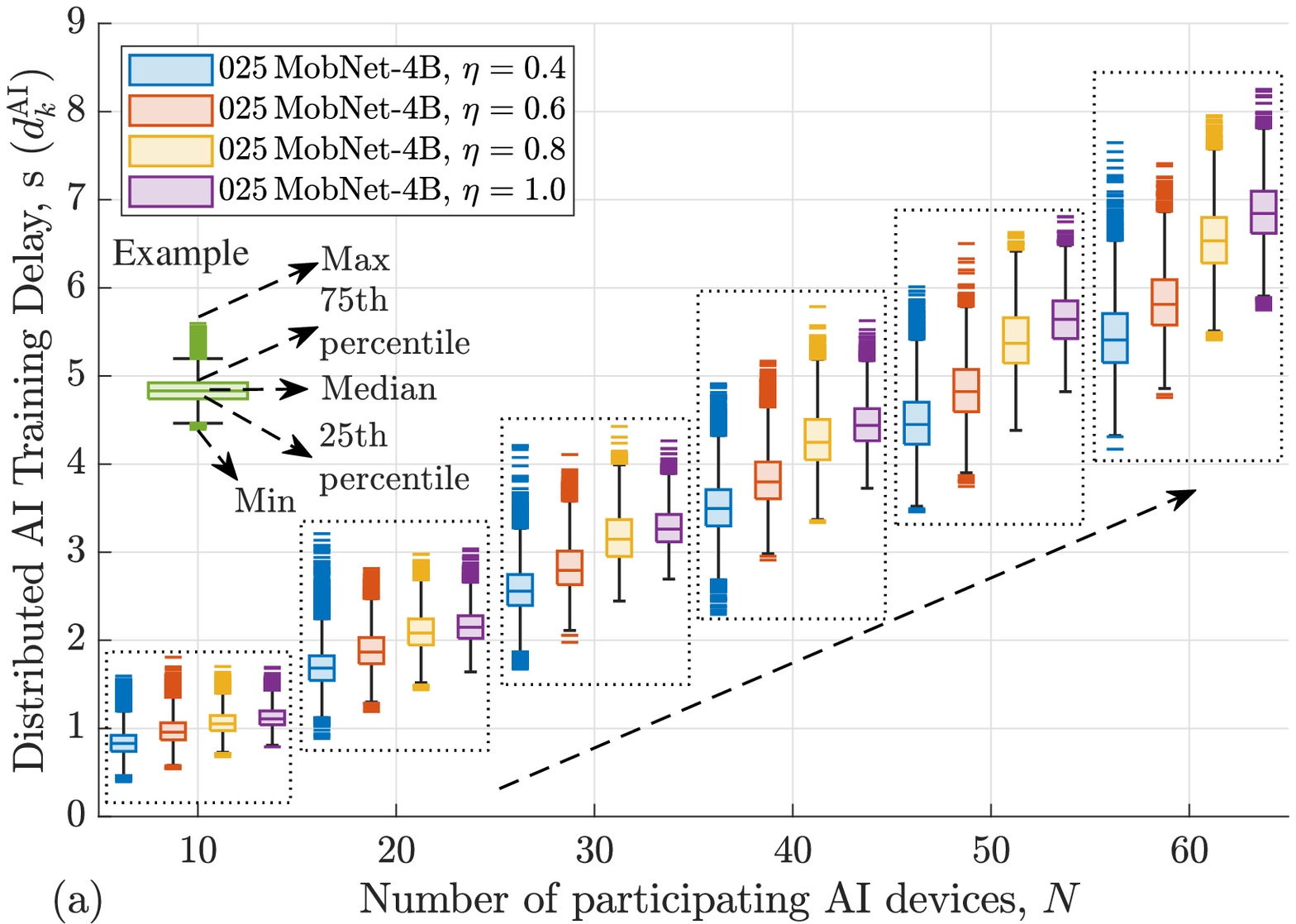}
		\vspace{-4.5mm}
    	%\vspace{-6.5mm}
    	%\caption{}
    	%\label{fig:delay025}
 %       \vspace{-1mm}
	\end{subfigure}
%  	\hspace{-4cm}
    %\vspace{-2mm}
	\begin{subfigure}[b]{0.5\textwidth}
    	\centering
    	\includegraphics[width=.99\columnwidth,keepaspectratio]{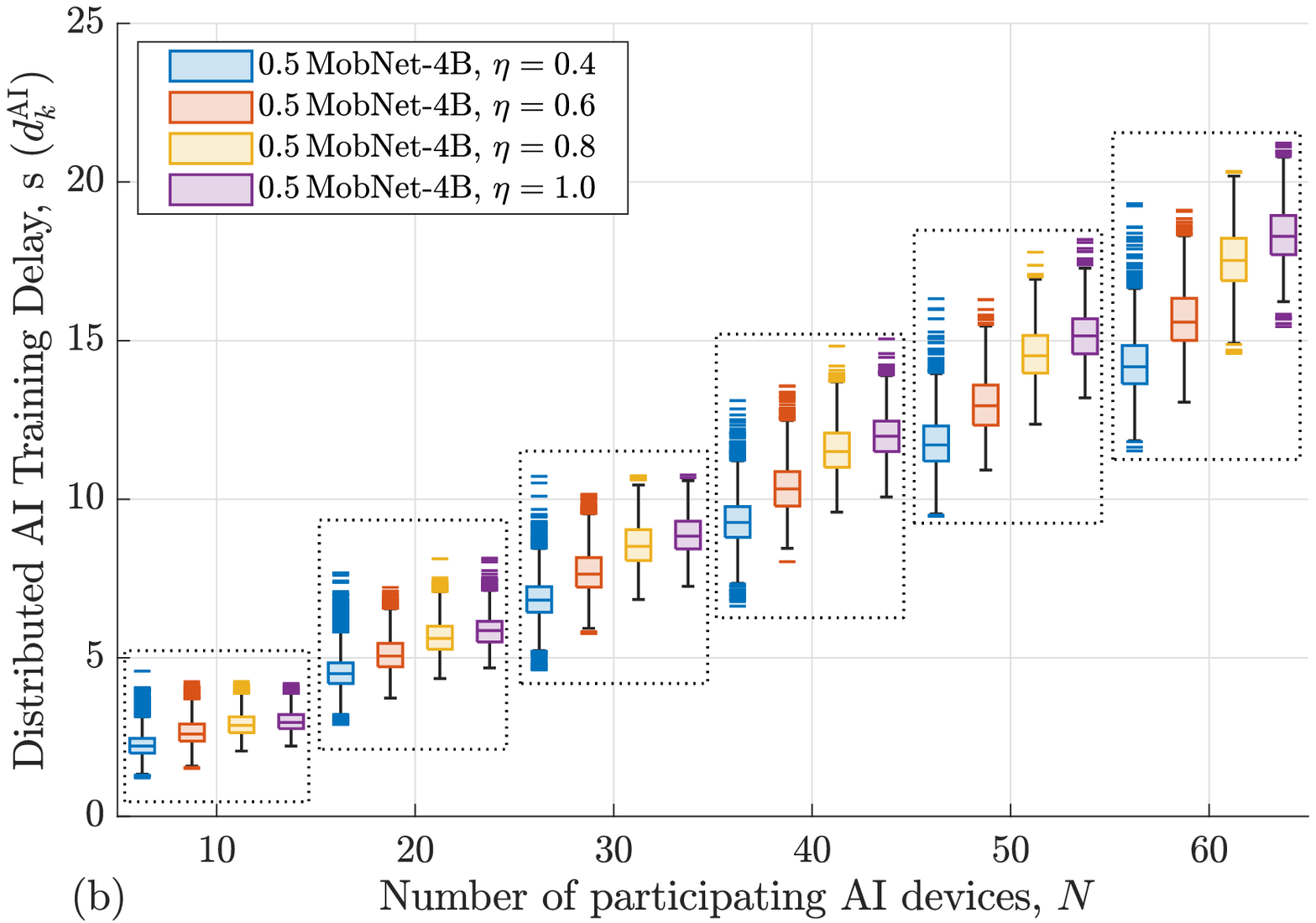}
    	\vspace{-4.5mm}
    	%\vspace{-6.5mm}
    	%\caption{}
    	%\label{fig:delay05}
	\end{subfigure}
	\hspace{3cm}
	\begin{subfigure}[b]{0.5\textwidth}
    	\centering
    	\includegraphics[width=.99\columnwidth,keepaspectratio]{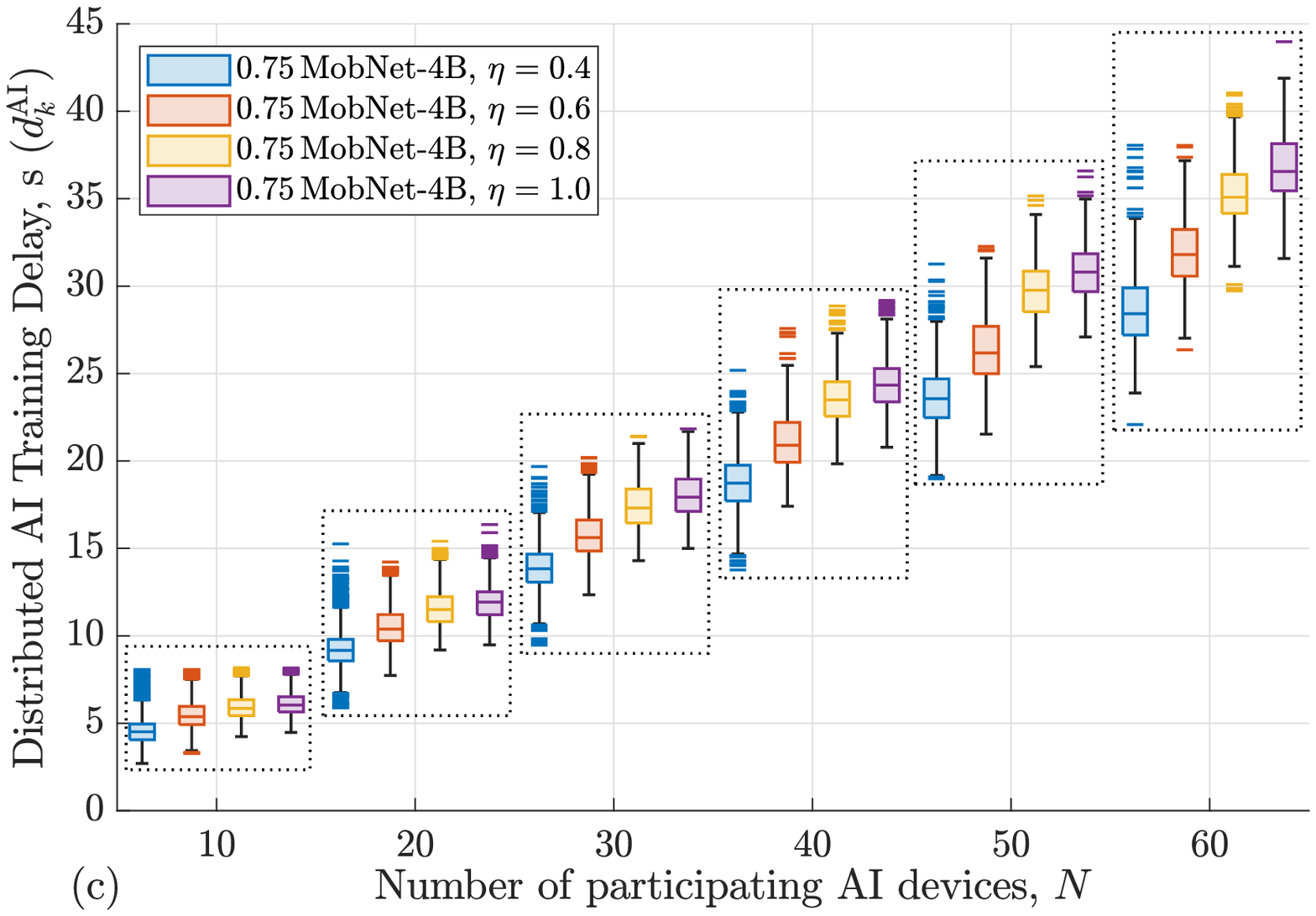}
        \vspace{-4.5mm}
    	%\vspace{-6.5mm}
    	%\caption{}
    	%\vspace{-2mm}
    	%\label{fig:delay075}
	\end{subfigure}
% 	\hspace{1cm}
	\begin{subfigure}[b]{0.5\textwidth}
    	\centering
    	\includegraphics[width=.99\columnwidth,keepaspectratio]{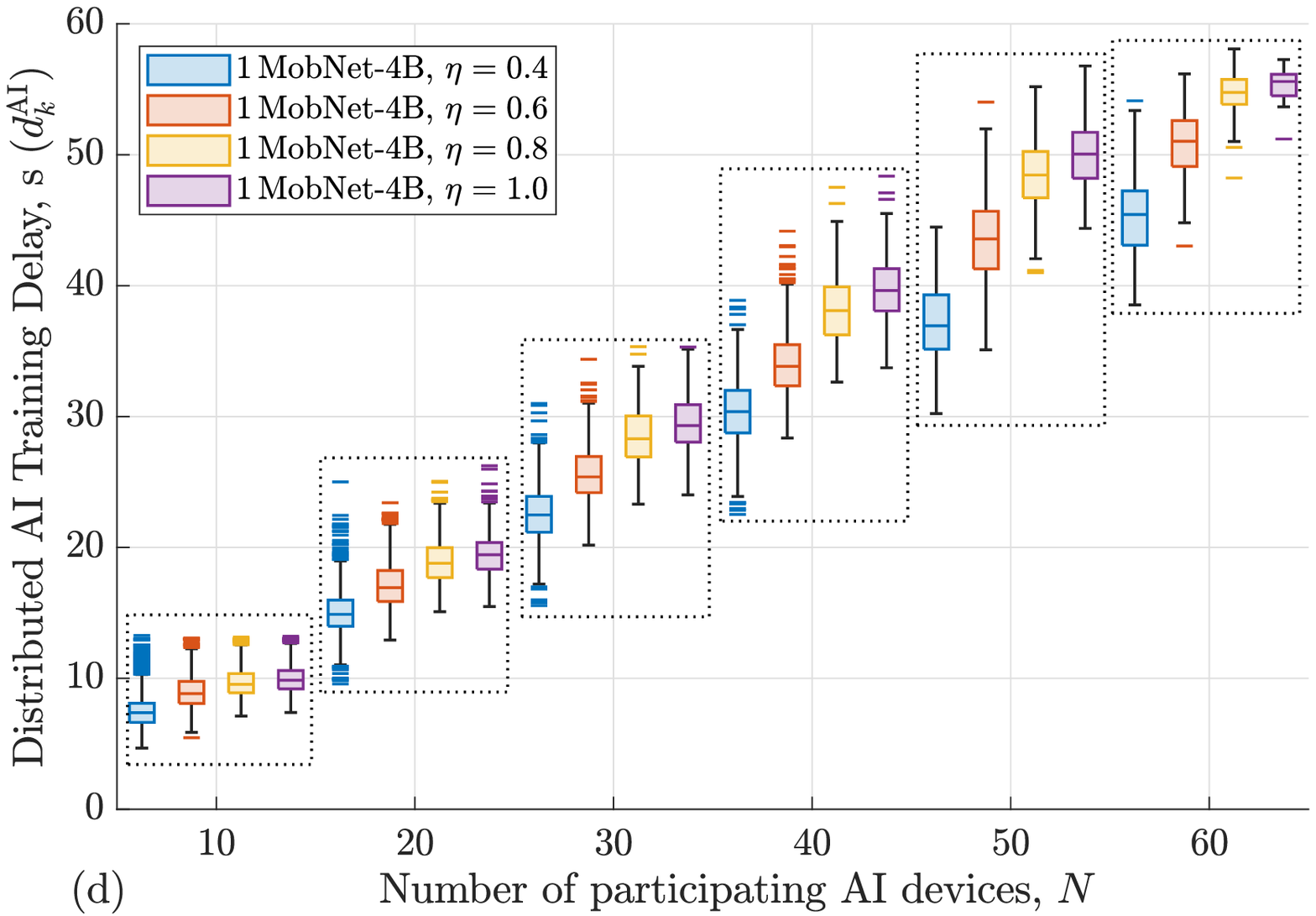}
		\vspace{-4.5mm}
    	%\vspace{-6.5mm}
    	%\caption{}
    	%\vspace{-2mm}
    	%\label{fig:delay1}
	   %	\vspace{-7mm}
	\end{subfigure}
	\caption{The distribution of the distributed AI training delay, $d_k^{\mathrm{AI}}$, for different $\eta$ and $N$. Each box plot represents the minimum, $25$th percentile, median, $75$th percentile, and maximum of the observed training delay samples per iteration (as shown via the example in (a) with green box) for a specific $\eta$ and $N$.}
	\label{fig:delay}
	\vspace{-5mm}
\end{figure*}

\figurename\,\ref{fig:delay} depicts the distribution of per-iteration delay of distributed AI, $d_k^{\mathrm{AI}}$ formulated in \eqref{eq:aiDelay}, for different $N$ and $\eta$ with $0.25$\,MobileNet (\figurename\,3a), $0.5$\,MobileNet (\figurename\,3b), $0.75$\,MobileNet (\figurename\,3c), and $1$\,MobileNet (\figurename\,3d). Each box represents the minimum, $25$th percentile, median, $75$th percentile and maximum observed training delay (refer to the example box in \figurename\,3a). In general, the distributed \gls{ai} training delay rises with the number of distributed \gls{ai} participants. 
% Although the model size in $1$\,MobileNet is $4$ times the model size in $0.25$\,MobileNet, its median training delay could be as high as $10$x. 

On \texttt{Eval1}, \figurename\,\ref{fig:delay} shows that higher $\eta$ results in higher training delay. As $\eta$ grows, the master node should wait for a larger portion of participating devices in which it is more likely to include stragglers in poor channel condition, thus leading to higher training delay. This higher training delay could explain the improvement in availability in \figurename\,\ref{fig:avail} for \texttt{Eval1} case. For example, \figurename\,3d shows that the median training delay for \texttt{Eval1}$\mathtt{\left(60{,}1\right)}$ increases $22\%$ from \texttt{Eval1}$\mathtt{\left(60{,}0.4\right)}$  to $55.6$\,s. Hence, it is much more likely for \gls{ai} devices in \texttt{Eval1}$\mathtt{\left(60{,}1\right)}$ to wait for the global model from master node, while the master node itself is waiting to receive the local models from stragglers. Such excessive pause in training leads to $2.7$\% improvement in $1$st percentile availability for \texttt{Eval1}$\mathtt{\left(60{,}1\right)}$ compared to \texttt{Eval1}$\mathtt{\left(60{,}0.4\right)}$ in \figurename\,3d. However, from convergence perspective, for a given $\calU$, increasing  $\eta$ results in a lower variance for the  gradient noise, potentially decreasing the number of iterations required to converge \cite{Bottou2018SIAM}.

For \texttt{Eval2}, our results in \figurename\,\ref{fig:avail} and \figurename\,\ref{fig:delay} demonstrate that using extra devices to leverage diversity, and thus alleviating the stragglers' problem, can lead to contradicting outcome (i.e., it could reduce the $1$st percentile availability by several percents and increase the training delay by tens of seconds). For example, in \figurename\,\ref{fig:avail1}, the \texttt{Eval2}$\mathtt{\left(50{,}0.6\right)}$'s $1$st percentile availability is down to $0.922$ from $0.937$ in \texttt{Eval2}$\mathtt{\left(30{,}1\right)}$. Not to mention \figurename\,3d, where the median of the training delay for \texttt{Eval2}$\mathtt{\left(50{,}0.6\right)}$ is $43.5$, $1.5$x the median training delay for \texttt{Eval2}$\mathtt{\left(30{,}1\right)}$. It seems that (i) high load, which is due to the transmission of large local and global models, and (ii) limited bandwidth on our system contributed to this contradictory outcome, thus overweighting the diversity gain.

The results in \cite{mobileNets} on the ImageNet dataset (a substantial visual database intended for use in the research on visual object recognition) show that $0.25$\,MobileNet, $0.5$\,MobileNet, $0.75$\,MobileNet, and $1$\,MobileNet could achieve an accuracy of $50.6\%$, $63.7\%$, $68.4\%$, and $70.6\%$, respectively. From \figurename\,\ref{fig:delay}, the higher accuracy of bigger models comes at the price of a higher distributed \gls{ai} training delay per global update iteration. This higher per-iteration complexity, together with the need for having more iterations for bigger models to converge result in a much higher convergence time. Furthermore, on the one hand, the accuracy of distributed \gls{ai} rises with the number of participating \gls{ai} devices. However, as the number of participating \gls{ai} devices continues to grow, the rate of such increase decelerates \cite{saadConvergence}. On the other hand, our results in \figurename\,\ref{fig:avail} and \figurename\,\ref{fig:delay} reveal that the number of participating \gls{ai} devices ($N$) heavily affects $\alpha_i$ and $d_k^{\mathrm{AI}}$. These trade-offs highlight the significance of efficient user selection in implementing distributed \gls{ai} techniques on future cellular networks.
% \begin{itemize}
%     \item The impact on \gls{urllc} (i.e., UL/DL Delay, \gls{sinr}, and availability)
%     \item \gls{ai} delay, convergence (time in x-axis, and test accuracy for several $k/n$ and different \gls{dnn} size)
%     \item can we get such conclusion: For a range of $k/n$ we have good convergence and for higher $k/n$ we show that we still have good convergence but the convergence time is not good (vs iteration is perhaps not changed much)
% \end{itemize}
%%%%%%%%%%%%%%%%%%%%%
% NEW SECTION
%%%%%%%%%%%%%%%%%%%%%
\section{Conclusions}\label{sec:conculsions}
In this paper, we studied the trade-offs between the availability of \gls{urllc} service and the convergence time of distributed training. We leveraged the already existing 5G-NR \gls{qos} handling to separate the traffic of the two services. Using our near-product \gls{3gpp}-compliant simulations, we showed that the introduction of \gls{ai} traffic has a considerable side effect on \gls{urllc} availability, which declines with the number of participating \gls{ai} devices. Although an increase in the number of \gls{ai} devices can improve the training accuracy of the distributed \gls{ai}, we showed that the cost could be an intolerable increase in training delay, thanks to the 5G-NR scheduler. 

As for future directions, we note that user-selection algorithms that jointly consider distributed \gls{ai}'s performance and the coexisting \gls{urllc} service are of crucial importance. Besides, new dynamic slicing approaches that optimize bandwidth allocation for \gls{ai} slice and \gls{urllc} slice can improve the coexistence performance.
\vspace{-1mm}
\section*{Acknowledgment}
The authors would like to sincerely thank Abdulrahman Alabbasi and Jonas Kronander for their insightful comments.
% If we properly set the hyperparameters of \gls{ai}, for a range of hyperparameters we can have negligible impact on \gls{urllc} while keeping the convergence time reasonably short

% use section* for acknowledgment
%\section*{Acknowledgment}

\ifCLASSOPTIONcaptionsoff
\newpage
\fi

%\begin{thebibliography}{1}
%	
%	\bibitem{IEEEhowto:kopka}
%	H.~Kopka and P.~W. Daly, \emph{A Guide to \LaTeX}, 3rd~ed.\hskip 1em plus
%	0.5em minus 0.4em\relax Harlow, England: Addison-Wesley, 1999.
%	
%\end{thebibliography}
\Urlmuskip=0mu plus 1mu\relax
\bibliographystyle{IEEEtran}
\vspace{-2mm}
% argument is your BibTeX string definitionsref and bibliography database(s)

\bibliography{./Components/Metafiles/ref}
\clearpage

\end{document}